\documentclass[11pt]{article}

\usepackage[totalheight = 23cm, totalwidth = 17cm]{geometry}
\usepackage{amssymb,amsmath,amsfonts,amsbsy,graphicx,bm,cancel,xcolor,ulem,hyperref}
\usepackage{arydshln,enumitem,multirow}

\hypersetup{colorlinks, bookmarksopen, bookmarksnumbered,
citecolor=teal, linkcolor=orange, pdfstartview=FitH, urlcolor=cyan}

\newcommand{\mpl}{m_{\rm Pl}}
\newcommand{\calJ}{{\cal J}}
\newcommand{\calK}{{\cal K}}
\newcommand{\calO}{{\cal O}}
\newcommand{\calP}{{\cal P}}
\newcommand{\calR}{{\cal R}}

\begin{document}

\begin{titlepage}

\rightline{\footnotesize{APCTP-Pre2026-009}}

\begin{center}

\vskip 3em

{\Large \bf 
Reheating as a variational probe of cosmological observables
}

\vskip 3em

{\large Jinn-Ouk Gong$^{a,b}$}

\vskip 0.5cm

{\it
$^{a}$Department of Science Education, Ewha Womans University, Seoul 03760, Korea
\\
$^{b}$Asia Pacific Center for Theoretical Physics, Pohang 37673, Korea
}

\end{center}

\vskip 1.2cm

\begin{abstract}

We formulate reheating as a constrained variational problem in the space of equation-of-state histories, rather than attempting to describe it through microscopic models. We introduce a regularized functional framework that identifies reheating histories which extremize a given cosmological observable under minimal physical assumptions. As illustrative applications, we consider prompt gravitational waves, induced gravitational waves, and primordial black holes. We find that different observables select qualitatively different regions of reheating-history space. These examples demonstrate that cosmological observables define distinct extremal directions in reheating-history space and can therefore be used to systematically explore the space of post-inflationary expansion histories.

\end{abstract}

\end{titlepage}

\renewcommand*{\thefootnote}{\arabic{footnote}}
\setcounter{footnote}{0}

\newpage

\section{Introduction}
\label{sec:intro}

While primordial inflation~\cite{Guth:1980zm,Linde:1981mu,Albrecht:1982wi} successfully explains the observed homogeneity of the universe and the origin of primordial fluctuations as the observations of the cosmic microwave background (CMB) have confirmed~\cite{Planck:2018vyg}, the energy in the inflation sector must be transferred to the usual matter contents with non-negative pressure to commence the hot big bang evolution. This process of energy transfer is called ``reheating'' (for reviews, see e.g.~\cite{Bassett:2005xm,Allahverdi:2010xz,Amin:2014eta}). The reheating epoch provides the essential bridge between inflation and the standard radiation-dominated universe. Although it does not directly affect the large-scale inflationary predictions in the simplest scenarios, the reheating history can leave observable imprints on a variety of cosmological probes. Well-known examples include the mapping between inflationary scales and CMB observables through the reheating expansion history~\cite{Martin:2014nya}, spectral distortions~\cite{Chluba:2013pya}, relic abundances~\cite{Giudice:2000ex}, gravitational waves (GWs)~\cite{Easther:2006gt}, and other probes of the early universe.

At a coarse-grained level, reheating may be characterized by the equation of state $w$\footnote{
The effective equation of state $w$ employed in this work should be viewed as a coarse-grained description of reheating. While such a description is convenient for characterizing the macroscopic expansion history, it need not correspond to the most fundamental microscopic description. Indeed, it has been argued that phenomena which appear non-perturbative at a macroscopic level may admit a perturbative microscopic description in terms of the underlying quantum degrees of freedom, see e.g.~\cite{Dvali:2022vzz}. Our analysis remains agnostic about the microscopic origin of the reheating dynamics and focuses instead on the effective expansion history relevant for cosmological observables.
}~\cite{Dai:2014jja,Cook:2015vqa}. The evolution of $w$ remains, however, largely unconstrained. This is partly because reheating process typically ends up with thermal equilibrium, and any model-dependent features are swept away~\cite{Harigaya:2013vwa,Mukaida:2015ria}. Moreover, to describe the detail of the energy transfer from the end of inflation, we must specify all the interaction channels between the inflation and visible sectors, which are completely behind the veil. Yet, $w$ is subject to robust and minimal constraints that the endpoint values are fixed by the end of inflation and the onset of radiation domination, with $w = -1/3$ and $1/3$ respectively~\cite{Choi:2025xye}. This observation suggests a natural variational interpretation of reheating: the function is free to vary in the interior of the interval, while its boundary values are fixed.

Unlike ordinary variational problems in classical mechanics, however, reheating does not come equipped with a unique equation of motion for $w$. This is in sharp contrast to the energy density $\rho$, the evolution of which is determined by a single universal equation -- the conservation equation. However, the evolution of $\rho$ is not determined by microphysics but is governed by symmetry, and that's why it obeys a universal conservation law~(see e.g. \cite{Mukhanov:2005sc,Weinberg:2008zzc}). By contrast, the pressure, or equivalently $w$, is determined by the properties of the fluid. But our ignorance on the first principles of the microphysics of reheating prohibits us from determining the functional form of $w$, so that the preferred evolution of $w$ as a function of, say, the number of $e$-folds $N$ is not uniquely determined.

A natural way to proceed is, therefore, to characterize reheating histories not by attempting to reconstruct the true dynamics, but by asking how they impact observable quantities -- which histories are preferred by a given physical observable. This shifts the focus from determining $w$ itself to understanding its observable consequences. 
In this sense, observables provide a natural organizing principle for the space of reheating histories. We adopt this operational viewpoint here. Instead of attempting to derive the reheating dynamics from microscopic physics, we characterize reheating histories by asking the following question: ``Which equation-of-state histories $w$, consistent with minimal physical assumptions, extremize a given observable?'' Different observables then define different extremal directions in the functional space of reheating histories. This formulation shifts the focus from reconstructing reheating to probing its observable imprints in a controlled and model-independent manner.

In this work we formulate reheating as a constrained functional optimization problem and introduce a variational framework for exploring the space of reheating histories. We construct a regularized functional, parameterize admissible histories subject to the endpoint conditions, and identify the histories that extremize a given observable. As illustrative applications, we consider prompt GWs, induced GWs~\cite{Ananda:2006af,Baumann:2007zm}, and primordial black holes (PBHs)~\cite{Zeldovich:1967lct,Hawking:1971ei,Carr:1974nx}. These examples demonstrate that different observables select qualitatively different regions of reheating-history space and reveal distinct physical features of the post-inflationary expansion history.

\section{Variational formulation for reheating}
\label{sec:variation}

Our goal is to therefore identify reheating histories that extremize a given observable $\calO[w]$ under the minimal physical constraints. In principle, this corresponds to a constrained functional optimization problem for $\calO[w]$. However, in the absence of detailed microphysical input, the space of admissible functions $w(N)$ is extremely large, and naive extremization of $\calO[w]$ alone can lead to unphysical or highly oscillatory solutions that exploit this freedom. In other words, the problem is ill-posed without additional criteria that select physically reasonable histories.
To address this, we introduce a regularized functional:
\begin{equation}
\label{eq:ext-J}
\calJ[w] \equiv \calO[w] - \lambda I_\text{prior}[w] \, .
\end{equation}

The role of the prior is to suppress unphysical or pathological reheating histories that would otherwise dominate the extremization due to the large functional freedom of $w(N)$. In the absence of a first-principles description of reheating, the choice of $I_\text{prior}[w]$ should be understood in the spirit of an effective theory~(see e.g.~\cite{Burgess:2007pt}): It encodes minimal and general expectations about the background evolution without committing to a specific microphysical model. In particular, we require that physically reasonable histories vary smoothly on the time scales relevant for the background expansion. This motivates the leading derivative term $I_\text{prior}[w] \supset \int dN\, (w')^2$, with a prime denoting a derivative with respect to $N$, which penalizes rapid variations and represents the lowest-order operator in a derivative expansion. We further allow for a soft anchoring of the evolution around a fiducial value $w_*$ through a quadratic term $(w - w_*)^2$, which prevents large excursions that are not supported by physical intuition, while remaining agnostic about the detailed dynamics. Taken together, we use the following prior:
\begin{equation}
I_\text{prior}[w] = \int_0^{N_\text{rh}} dN \Big[ (w')^2 + \alpha \big( w - w_* \big)^2 \Big] \, .
\end{equation}
This construction provides a minimal, model-independent prior that renders the extremization problem well-defined, while preserving sensitivity to physically relevant variations in the reheating history.

To make the variational problem concrete, we parametrize the reheating history as a deformation around a fixed reference profile:
\begin{equation}
\label{eq:w-profile}
w(N) = w_\text{ref}(N) + \sum_{k=1}^K \frac{b_k}{k^p} \phi_k(N) \, ,
\end{equation}
where $N$ is the number of $e$-folds during reheating, satisfying $0 \leq N \leq N_\text{rh}$, and the coefficients $\{b_k\}$ encode deviations from the reference evolution.

\begin{itemize}

\item The reference profile $w_\text{ref}(N)$ is chosen to be the simplest interpolation consistent with the boundary conditions:
\begin{equation}
w_\text{ref}(N) = - \frac{1}{3} + \frac{2}{3} \frac{N}{N_\text{reh}} \, ,
\end{equation}
which provides a smooth baseline connecting the end of inflation and the beginning of radiation domination without introducing additional structure.

\item The basis functions $\phi_k(N)$ are taken to vanish at the endpoints, ensuring that all variations automatically preserve the boundary conditions. A convenient and complete choice is
\begin{equation}
\phi_k(N) = \sin \bigg( \frac{k\pi N}{N_\text{rh}} \bigg) \, ,
\end{equation}
which forms an orthogonal set on the interval $[0, N_\text{rh}]$ and provides a systematic decomposition of admissible reheating histories.

\item The factor $k^{-p}$ with $p$ being a number of $\calO(1)$ suppresses higher-$k$ modes, implementing a mild form of spectral regularization that complements the prior functional. This ensures that the expansion remains well-behaved and that the extremization is not dominated by highly oscillatory configurations.

\end{itemize}

This parametrization provides a systematically improvable representation of smooth reheating histories consistent with the endpoint constraints. In practice, the expansion converts the functional extremization problem into a finite-dimensional optimization over the coefficients $\{b_k\}$, which are varied numerically to extremize $\calJ[w]$. In the limit $K \to \infty$, this expansion can approximate any sufficiently smooth function satisfying the boundary conditions. In the following, we apply this framework to several observables and examine which reheating histories are selected by their extremization.

\section{Applications to physical observables}
\label{sec:observables}

We now apply the variational framework developed in the previous section to several observables associated with reheating. The purpose of the following examples is not to identify the physically realized reheating history, but to understand which features of reheating are selected by different observables. For each observable, we extremize \eqref{eq:ext-J} over the space of reheating histories satisfying the fixed endpoint conditions $w(0) = -1/3$ and $w(N_\text{rh}) = 1/3$. Apart from the physical requirement that reheating should not revert to inflation, i.e. $w<-1/3$ is excluded, the space of admissible histories is left unrestricted. The resulting histories reveal which features of the reheating background are preferred by the observable itself. We emphasize that the purpose of the following examples is not to provide precise predictions for the corresponding observables, but to illustrate how different observables select different regions of reheating-history space.

In all examples we use \eqref{eq:w-profile} and optimize over the coefficients $\{b_k\}$, with $K=12$, $p=1$ and $w_* = 0$, and we adopt $N_\text{rh} = 6$ as a representative value. For observables involving primordial scalar power spectrum $\calP_\calR(k)$, i.e. induced GWs and PBHs, we adopt a phenomenological scalar spectrum consisting of a nearly scale-invariant component and a Gaussian bump on small scales, with the peak amplitude $A_\text{peak} = 0.02$ and width $\sigma_k = 0.5$. The spectrum parameters are fixed throughout this work. To assess the robustness of the results, we repeat the extremization for several values of the prior parameters $(\lambda,\alpha)$, which control the overall strength and shape of the regularization. Unless otherwise stated, the qualitative behavior is insensitive to the particular choice of prior. In practice, we focus on values of $(\lambda,\alpha)$ for which the extremized histories are stable under changes of the optimization seed and remain qualitatively consistent across the tested parameter range.

\subsection{Prompt GWs}

We first consider prompt GWs produced during reheating. This provides a clean first example of the variational framework, since the observable is directly sourced by the background evolution during reheating. Most studies of reheating GWs focus on specific microscopic production mechanisms, such as inflaton decay~\cite{Nakayama:2008wy}, scattering~~\cite{Choi:2024ilx}, or preheating dynamics~\cite{Easther:2006gt,Khlebnikov:1997di}. Here we adopt a complementary viewpoint and isolate the role of the background expansion history, asking which reheating histories extremize the final GW abundance independent of the underlying microscopic realization. The observable is defined from the GW energy density generated by inflaton scattering~\cite{Choi:2024ilx}, which can be written as
\begin{equation}
\label{eq:OpGW}
\calO_{\rm pGW}[w]
\equiv
\rho_{\rm GW}[w]
=
\int_0^{N_{\rm rh}} dN\,
e^{-4(N_{\rm rh}-N)}
\frac{(1+w)\Gamma_h \rho_\phi}{H}
\, ,
\end{equation}
where
\begin{equation}
\Gamma_h = \frac{\rho_\phi m_\phi}{32\pi\mpl^4} \, .
\end{equation}
The dependence on the reheating history enters through $w(N)$, the background inflaton energy density $\rho_\phi(N)$, and the Hubble rate $H(N)$.

\begin{figure}[h]
\begin{center}
\includegraphics[width=1.0\textwidth]{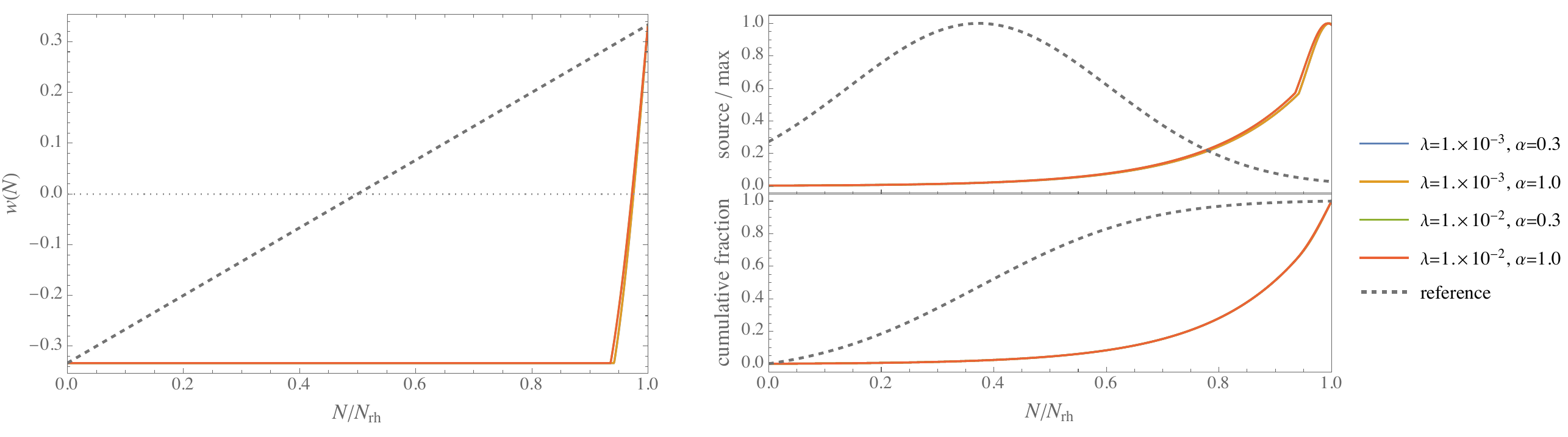}
\end{center}
\vspace{-1.5em}
\caption{Extremized reheating histories for prompt GW production. The left panel shows the optimized equation-of-state histories $w(N)$, while the right panel displays the normalized prompt-GW source contribution (upper) and cumulative GW abundance (lower). Coloured curves correspond to representative choices of $(\lambda,\alpha)$, while the dashed curve denotes the linear reference history. All optimized solutions converge to a nearly identical reheating history characterized by a prolonged period near $w\approx-1/3$. The source contribution remains concentrated near the end of reheating, where most of the GW abundance is generated.
}
\label{fig:pGW}
\end{figure}

We extremize \eqref{eq:ext-J} with the observable functional \eqref{eq:OpGW} using the expansion \eqref{eq:w-profile} with $K=12$. We consider three values of the prior strength parameter, $\lambda = \{10^{-3},3\times10^{-3},10^{-2}\}$, and two representative values of the anchoring parameter, $\alpha = \{0.3,1.0\}$. In all cases the optimization converges to essentially the same reheating history, with variations below the visual resolution of Figure~\ref{fig:pGW}. The corresponding solutions yield $\calO_\text{pGW} \approx 1.9 \times 10^{-10}$, $N_\text{eMD}\approx0.07$, and $N_\text{max}\approx5.97$. Here, $N_\text{eMD} \equiv \int_0^{N_{\rm rh}} dN\, \big[1-3w(N)\big]/2$ measures the integrated matter-like weighting rather than the literal duration of a matter-dominated phase, and $N_\text{max}$ denotes the $e$-fold at which the prompt-GW source reaches its maximum. The extremized histories therefore exhibit only a very short effective matter-like phase and remain close to $w=-1/3$ throughout most of reheating.

This result can be understood directly from the structure of \eqref{eq:OpGW}. The prompt-GW production rate scales as $\Gamma_h\rho_\phi \sim \rho_\phi^2$, favouring reheating histories that preserve a large inflaton energy density for as long as possible. Since $\rho_\phi \propto \exp \big[ -3(1+w)N \big]$ during reheating, the slowest dilution is achieved near the lower allowed boundary $w=-1/3$. The variational procedure therefore selects histories that remain close to this limit throughout most of reheating and approach radiation domination only near $N_\text{rh}$. Figure~\ref{fig:pGW} shows that the dominant source contribution is concentrated near the end of reheating, where the cumulative GW abundance rises rapidly.

\subsection{Induced GWs}

Compared to prompt GWs, induced GWs provide a considerably more non-trivial probe of reheating histories. As mentioned, here and in the next subsection we adopt a phenomenological shape for $\calP_\calR(k)$ with a Gaussian bump. The induced GW energy density is schematically written as~\cite{Kohri:2018awv}
\begin{equation}
\calO_\text{iGW}[w]
\equiv 
\Omega_\text{iGW}[w] 
= 
\int du dv \int dx \, \calK_\text{KT}(u,v,x) \calP_\calR(uk) \calP_\calR(vk) \, ,
\end{equation}
where $u$ and $v$ parametrize the scalar momenta in the convolution, while $x\equiv k\eta$ tracks the time evolution of the tensor source. The reheating history enters through the background-dependent relation between $N$, $x(N)$, and the horizon-reentry scale $k_\text{re}(N)$, and hence through the way the kernel is sampled during reheating. In contrast to the prompt GW case, the observable is therefore intrinsically non-local in the reheating background and depends on the detailed temporal structure of the source evolution.

\begin{figure}[h]
\begin{center}
\includegraphics[width=0.45\textwidth]{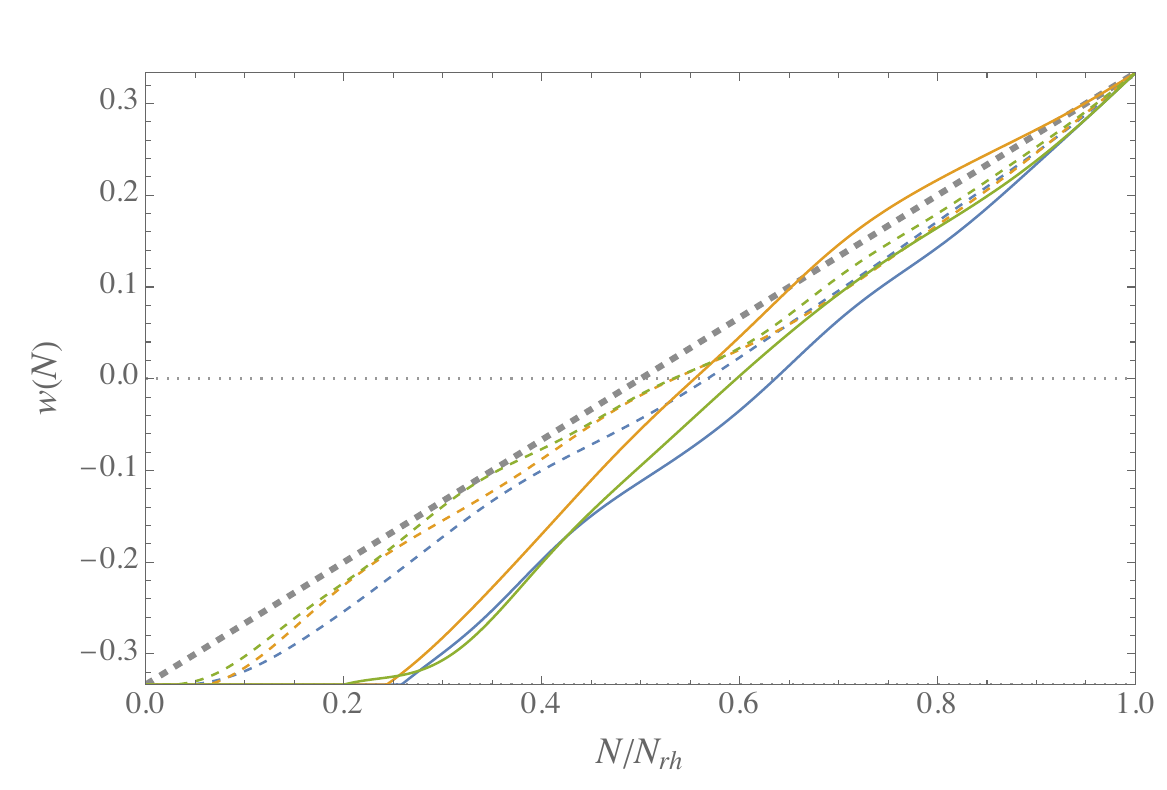}
\includegraphics[width=0.54\textwidth]{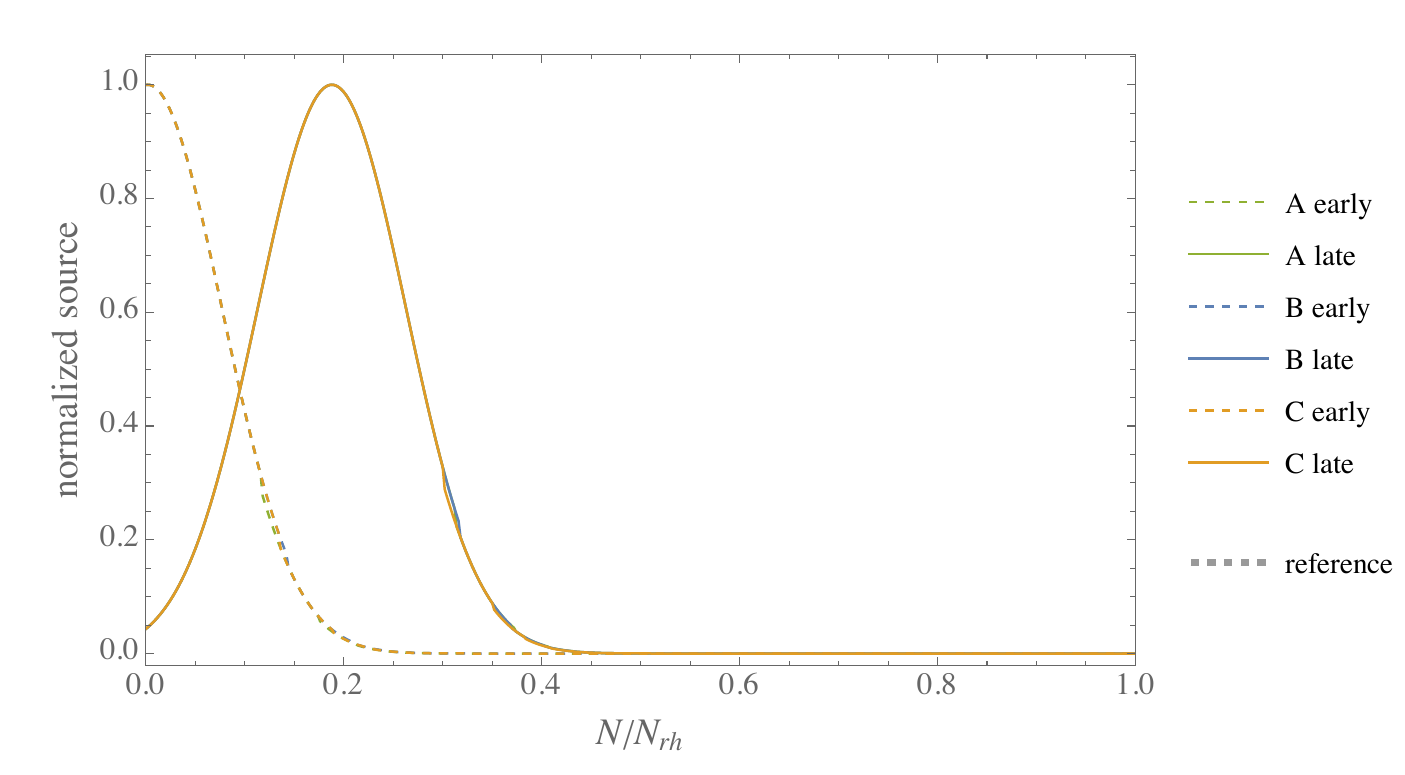}
\end{center}
\vspace{-1.5em}
\caption{Extremized reheating histories for induced GW production. The left panel shows the optimized equation-of-state histories $w(N)$, while the right panel displays the effective source contribution entering the induced-GW observable. Coloured curves correspond to representative choices of  $(\lambda,\alpha)$, and the thick dashed curve denotes the linear reference history. In contrast to prompt GWs, the extremized solutions develop an extended matter-like phase and a source contribution localized at a later time during reheating.}
\label{fig:iGW}
\end{figure}

We extremize the functional \eqref{eq:ext-J} using \eqref{eq:w-profile} with $K=12$. For all tested values of $(\lambda,\alpha)$, the optimization converges to the same qualitative family of reheating histories characterized by an extended matter-like phase. Representative solutions are shown in Figure~\ref{fig:iGW}, while the corresponding diagnostics are summarized in Table~\ref{table:iGW}.

\begin{table}[h]
\begin{center}
\begin{tabular}{c|c|c|c|c|c}
Branch type & $\lambda$ & $\alpha$ & $\calO_\text{iGW}$ & $N_\text{eMD}$ & $N_\text{max}$
\\
\hline\hline
A-early & \multirow{2}{*}{$5.0\times10^{-4}$} & \multirow{2}{*}{$0.10$} & $3.44\times10^{-4}$ & $3.32$ & $0.01$ 
\\
A-late &  & & $6.71\times10^{-4}$ & $3.79$ & $1.13$ 
\\\hline
B-early & \multirow{2}{*}{$5.0\times10^{-4}$} & \multirow{2}{*}{$0.05$} & $3.44\times10^{-4}$ & $3.24$ & $0.01$ 
\\
B-late & & & $6.69\times10^{-4}$ & $3.45$ & $1.13$ 
\\\hline
C-early & \multirow{2}{*}{$1.0\times10^{-4}$} & \multirow{2}{*}{$0.10$} & $3.42\times10^{-4}$ & $3.17$ & $0.01$ 
\\
C-late & & & $6.70\times10^{-4}$ & $3.70$ & $1.13$ 
\\\hdashline
Late & $1.0\times10^{-4}$ & $0.03$ & $6.77\times10^{-4}$ & $4.81$ & $2.87$ 
\end{tabular}
\end{center}
\caption{Representative diagnostics for the induced-GW branches shown in Figure~\ref{fig:iGW}.
The first six rows display paired early- and late-source solutions for three representative $(\lambda,\alpha)$ choices.
The last row shows an additional late-source example with substantially larger $N_\text{max}$, illustrating that the source localization is not restricted to a specific preferred time during reheating.}
\label{table:iGW}
\end{table}

The extremized histories shown in Figure~\ref{fig:iGW} exhibit a markedly different behavior from the prompt-GW case. The effective matter-dominated duration increases to $N_\text{eMD} \gtrsim 3$, indicating that induced GW production favours an extended matter-like reheating phase. A more detailed inspection of the optimized solutions reveals that the effective matter-dominated duration alone does not completely determine the final value of $\calO_\text{iGW}$. Among histories with comparable $N_\text{eMD}$, we find that larger values of $\calO_\text{iGW}$ are systematically associated with a stronger localization of the source contribution toward later times. In other words, even when two reheating histories generate a similar matter-dominated interval, the history whose dominant source term remains active until a later stage tends to produce a larger integrated GW signal. This is in agreement with~\cite{Inomata:2019ivs}.

This behaviour can be understood directly from the structure of the induced-GW kernel. Unlike prompt GW production, which depends directly on the instantaneous inflaton energy density, induced GWs are sourced by scalar perturbations whose contribution accumulates over an extended period. A longer matter-dominated epoch and late-time contributions together enhance the relevant source term while delaying its dilution, thereby increasing the final GW abundance. The variational procedure thus favors not only a prolonged matter-dominated phase but also a delayed concentration of the dominant source contribution. The induced-GW optimum is therefore characterized by two complementary features: long matter domination and late source localization.

\subsection{PBHs}

PBHs provide a complementary probe of reheating, because the abundance is exponentially sensitive to the collapse probability. For definiteness, we consider three representative prescriptions: the analytic threshold criterion (HYK)~\cite{Harada:2013epa}, the refined threshold criterion (EGS)~\cite{Escriva:2019phb}, and the peak-theory (PT) estimate~\cite{Yoo:2019pma,Yoo:2020dkz}. For each prescription $X$, we define the integrated abundance proxy
\begin{equation}
\label{eq:PBHobs}
\calO_\text{PBH}^{(X)}[w] 
\equiv 
\log\beta_X[w]
=
\log \bigg\{
\int d\log{k} \, \calK_X(k;w) \, \exp \big[ -E_X(k;w) \big]
\bigg\} 
\, ,
\end{equation}
where $\calK_X$ denotes the prescription-dependent kernel weight and $E_X$ is the corresponding collapse exponent. The reheating history enters through the variance, the effective equation of state at horizon reentry, and the collapse threshold.

\begin{figure}[h]
\begin{center}
\includegraphics[width=\textwidth]{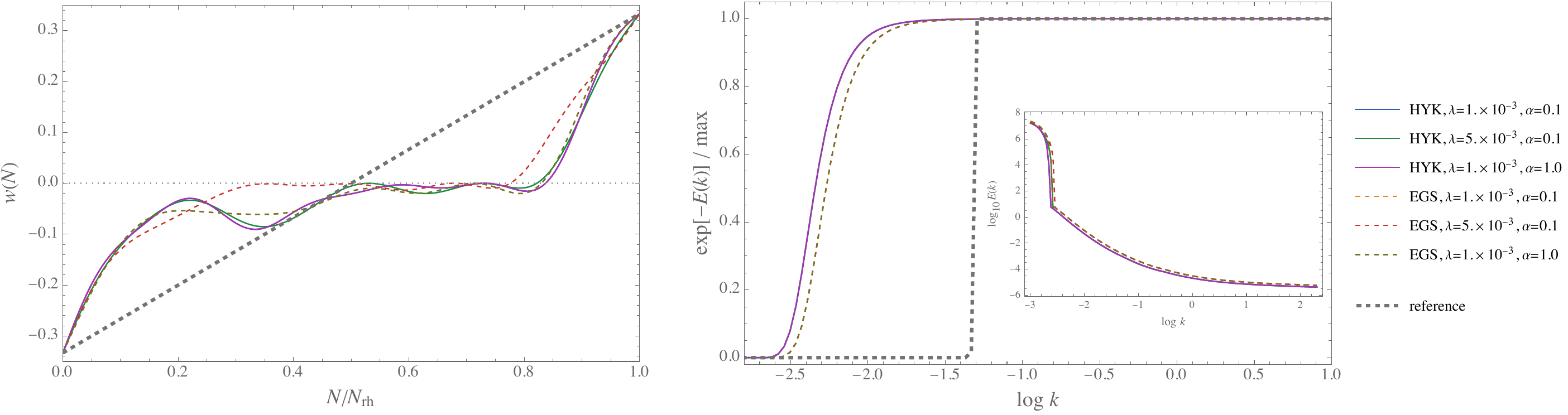}
\end{center}
\caption{Extremized reheating histories and PBH collapse weights obtained from the variational procedure. The left panel shows the optimized equation-of-state histories $w(N)$ for representative choices of $(\lambda,\alpha)$ using the HYK and EGS collapse prescriptions, while the right panel displays the corresponding normalized collapse weights $e^{-E(k)}$. The inset shows the collapse exponent $E(k)$ on a logarithmic scale. Despite variations in the prior parameters and collapse prescriptions, all optimized solutions converge to nearly identical histories characterized by an extended matter-dominated phase $N_\text{eMD} \sim 3$ and an early turn-on of efficient collapse $\log_{10}k_\text{turn} \sim -2.3$. The dashed curve denotes the linear reference history.}
\label{fig:PBH}
\end{figure}

We extremize \eqref{eq:ext-J} using \eqref{eq:w-profile} with $K=12$ for several values of the prior parameters $(\lambda,\alpha)$. Representative extremized solutions are shown in Figure~\ref{fig:PBH}, while the corresponding diagnostics are summarized in Table~\ref{table:PBH}. For the HYK prescription, all tested values of $(\lambda,\alpha)$ converge to nearly identical reheating histories. The resulting solutions are characterized by an extended matter-dominated phase $N_\text{eMD} \approx 3.15$ and a collapse-weight turn-on scale around $\log_{10}k_\text{turn} \approx -2.35$. Varying the prior parameters changes neither the integrated abundance nor the characteristic scales at a visible level. The same behaviour is found for the EGS prescription, which yields almost identical optimized histories and diagnostics.

\begin{table}[h]
\begin{center}
\begin{tabular}{c|c|c|c|c|c|c}
Prescription & $\lambda$ & $\alpha$ & $\calO_\text{PBH}$ & $N_\text{eMD}$ & $\log_{10}k_\text{turn}$ & $w_\text{turn}$ 
\\
\hline\hline
\multirow{3}{*}{HYK} & $10^{-3}$ & $0.1$ & $1.5332$ & $3.154$ & $-2.352$ & $-0.0000$ \\
& $5\times10^{-3}$ & $0.1$ & $1.5331$ & $3.164$ & $-2.352$ & $-0.0008$ \\
& $10^{-3}$ & $1.0$ & $1.5332$ & $3.154$ & $-2.352$ & $-0.0000$ \\
\hline
\multirow{3}{*}{EGS} & $10^{-3}$ & $0.1$ & $1.5173$ & $3.155$ & $-2.277$ & $0.0001$ \\
& $5\times10^{-3}$ & $0.1$ & $1.5174$ & $3.005$ & $-2.277$ & $0.0001$ \\
& $10^{-3}$ & $1.0$ & $1.5173$ & $3.155$ & $-2.277$ & $0.0001$ \\
\end{tabular}
\end{center}
\caption{Representative diagnostics of the extremized PBH solutions shown in Figure~\ref{fig:PBH}. For each collapse prescription and prior choice $(\lambda,\alpha)$, we list the integrated abundance proxy $\calO_\text{PBH}=\log\beta$, the effective matter-dominated duration $N_\text{eMD}$, the turn-on scale $k_\text{turn}$, and the corresponding local equation-of-state parameter $w_{\rm turn}$. The optimized solutions exhibit remarkable stability under variations of the prior parameters. HYK and EGS lead to nearly identical extremized histories and diagnostics, indicating that the variational preference for a long matter-dominated phase and an early collapse turn-on is largely independent of the collapse prescription. The PT prescription initially appears different due to a prescription-dependent peak-theory prefactor. Isolating this normalization contribution in a diagnostic optimization reveals that the collapse-exponent extremization converges to the same qualitative reheating history preferred by the HYK and EGS prescriptions.}
\label{table:PBH}
\end{table}

The left panel of Figure~\ref{fig:PBH} shows the extremized reheating histories, while the right panel displays the corresponding collapse weights $e^{-E(k)}$. Despite small differences in the detailed oscillatory structure of $w(N)$, the collapse weights remain nearly indistinguishable. The inset shows the collapse exponent $E(k)$, demonstrating that efficient collapse begins at essentially the same scale for all optimized solutions. This indicates that the variational preference is highly robust and does not depend sensitively on the choice of prior.

The physical origin of this behaviour can be understood from the structure of \eqref{eq:PBHobs}. Since the integrand is exponentially weighted by $e^{-E(k)}$, PBH production is controlled primarily by the scales at which the collapse exponent first becomes sufficiently small. Histories that shift the onset of efficient collapse to earlier scales receive a larger integrated contribution. At the same time, a prolonged matter-dominated phase enhances the collapse probability over a broad range of scales. The variational procedure therefore selects reheating histories that simultaneously maximize the effective matter-dominated duration and trigger efficient collapse as early as possible.

We have also repeated the analysis using the EGS and PT prescriptions. HYK and EGS lead to almost identical optimized histories and collapse-weight profiles. Although the PT prescription initially appears to prefer a different optimum, this behaviour can be traced to the overall normalization associated with the peak-theory counting factor. When the collapse exponent is extremized separately from this normalization contribution, the preferred reheating history converges to the same long-matter-dominated, early-collapse solution found in the HYK and EGS prescriptions. The exponent diagnostics are themselves nearly identical across all three collapse prescriptions, suggesting that the variational preference for long matter domination and early collapse onset is controlled by a common collapse exponent, and is therefore largely prescription independent.

We further checked the stability of the extremized solution under the variation of the prior parameters, including a reversal of the sign of $\alpha$. Remarkably, the extremized solution remains essentially unchanged. Since the prior functional merely plays the role of a coarse-grained regularization, this insensitivity indicates that the extremized history is selected by the structure of the PBH observable itself rather than by the prior. The long-matter-dominated, early-collapse solution therefore appears to define a unique variational attractor of PBH production.

Compared with the induced-GW case, the extremized PBH histories exhibit a similar preference for prolonged matter domination, but differ in the localization of the relevant contribution. While induced GWs favour a late-time concentration of the dominant source term, PBH formation is optimized by shifting the onset of efficient collapse toward earlier scales. In this sense, the two observables probe different aspects of the reheating history despite sharing a common preference for an extended matter-dominated epoch.

\section{Conclusions}
\label{sec:conc}

Reheating is usually described by a small number of effective parameters, such as an averaged equation of state or reheating duration. However, the background evolution during reheating is in general a continuous function $w(N)$, and different observables may probe different aspects of this history. In this work we introduced a variational framework that treats the reheating history itself as the object of optimization and asks which histories are preferred by a given observable.

To make the problem well posed, we introduced a regularized functional $\calJ[w]$ and parametrized reheating histories through a systematic basis expansion satisfying the endpoint conditions. The resulting problem reduces to a finite-dimensional optimization over the expansion coefficients while retaining the interpretation of a functional extremization in the space of reheating histories.

As applications of the variational framework, we considered three representative observables:

\begin{itemize}

\item Prompt GWs: The extremized histories remain close to the lower boundary $w=-1/3$ throughout most of reheating. This reflects the direct dependence of the production rate on the inflaton energy density and shows that the variational procedure selects histories that minimize dilution.

\item Induced GWs: The extremization favours an extended matter-like phase, indicating that the observable is sensitive not only to the instantaneous energy density but also to the temporal structure of the source evolution encoded in the induced-GW kernel. Distinct source-localization patterns can emerge, while sharing a common preference for prolonged matter-like evolution.

\item PBHs: The extremization selects histories with a prolonged matter-dominated phase and an early onset of efficient collapse. This indicates that the observable is controlled primarily by the scales at which the collapse exponent first becomes sufficiently small, rather than by the detailed shape of the subsequent evolution. The variational procedure therefore favours histories that shift efficient collapse to earlier scales while maintaining an extended matter-dominated phase.

\end{itemize}

More broadly, the variational perspective advocated here shifts the focus from predicting observables for a chosen reheating model to inferring which classes of reheating histories are selected by the observables themselves. While we have considered prompt GWs, induced GWs, and PBHs as illustrative examples, the framework is readily applicable to other probes of the early universe. Future applications could include spectral distortions, small-scale non-Gaussianity, isocurvature perturbations, or other observables sensitive to the post-inflationary expansion history. In this sense, cosmological observables can be viewed as probes of the geometry of reheating-history space, each selecting a distinct extremal direction. The variational framework developed here provides a systematic way to identify and compare these directions without committing to a specific microscopic model.

\section*{Acknowledgements}

I thank Gia Dvali, Jai-chan Hwang, Seong Chan Park and Alexander Vikman for helpful discussions.
This work is supported in part by Basic Science Research Program through the National Research Foundation of Korea (RS-2024-00336507) and the Ewha Womans University Research Grant of 2026 (1-2026-0486-001-1).
I am also grateful to the Asia Pacific Center for Theoretical Physics for hospitality while this work was under progress.


\begin{thebibliography}{99}




\bibitem{Guth:1980zm}
A.~H.~Guth,
Phys. Rev. D \textbf{23}, 347-356 (1981).




\bibitem{Linde:1981mu}
A.~D.~Linde,
Phys. Lett. B \textbf{108}, 389-393 (1982).




\bibitem{Albrecht:1982wi}
A.~Albrecht and P.~J.~Steinhardt,
Phys. Rev. Lett. \textbf{48}, 1220-1223 (1982).




\bibitem{Planck:2018vyg}
N.~Aghanim \textit{et al.} [Planck],
Astron. Astrophys. \textbf{641}, A6 (2020)
[erratum: Astron. Astrophys. \textbf{652}, C4 (2021)]
[arXiv:1807.06209 [astro-ph.CO]].




\bibitem{Bassett:2005xm}
B.~A.~Bassett, S.~Tsujikawa and D.~Wands,
Rev. Mod. Phys. \textbf{78}, 537-589 (2006)
[arXiv:astro-ph/0507632 [astro-ph]].




\bibitem{Allahverdi:2010xz}
R.~Allahverdi, R.~Brandenberger, F.~Y.~Cyr-Racine and A.~Mazumdar,
Ann. Rev. Nucl. Part. Sci. \textbf{60}, 27-51 (2010)
[arXiv:1001.2600 [hep-th]].




\bibitem{Amin:2014eta}
M.~A.~Amin, M.~P.~Hertzberg, D.~I.~Kaiser and J.~Karouby,
Int. J. Mod. Phys. D \textbf{24}, 1530003 (2014)
[arXiv:1410.3808 [hep-ph]].




\bibitem{Martin:2014nya}
J.~Martin, C.~Ringeval and V.~Vennin,
Phys. Rev. Lett. \textbf{114}, no.8, 081303 (2015)
[arXiv:1410.7958 [astro-ph.CO]].




\bibitem{Chluba:2013pya}
J.~Chluba and D.~Jeong,
Mon. Not. Roy. Astron. Soc. \textbf{438}, no.3, 2065-2082 (2014)
[arXiv:1306.5751 [astro-ph.CO]].




\bibitem{Giudice:2000ex}
G.~F.~Giudice, E.~W.~Kolb and A.~Riotto,
Phys. Rev. D \textbf{64}, 023508 (2001)
[arXiv:hep-ph/0005123 [hep-ph]].




\bibitem{Easther:2006gt}
R.~Easther and E.~A.~Lim,
JCAP \textbf{04}, 010 (2006)
[arXiv:astro-ph/0601617 [astro-ph]].




\bibitem{Dvali:2022vzz}
G.~Dvali and L.~Eisemann,
Phys. Rev. D \textbf{106}, no.12, 125019 (2022)
[arXiv:2211.02618 [hep-th]].




\bibitem{Dai:2014jja}
L.~Dai, M.~Kamionkowski and J.~Wang,
Phys. Rev. Lett. \textbf{113}, 041302 (2014)
[arXiv:1404.6704 [astro-ph.CO]].




\bibitem{Cook:2015vqa}
J.~L.~Cook, E.~Dimastrogiovanni, D.~A.~Easson and L.~M.~Krauss,
JCAP \textbf{04}, 047 (2015)
[arXiv:1502.04673 [astro-ph.CO]].




\bibitem{Harigaya:2013vwa}
K.~Harigaya and K.~Mukaida,
JHEP \textbf{05}, 006 (2014)
[arXiv:1312.3097 [hep-ph]].




\bibitem{Mukaida:2015ria}
K.~Mukaida and M.~Yamada,
JCAP \textbf{02}, 003 (2016)
[arXiv:1506.07661 [hep-ph]].




\bibitem{Mukhanov:2005sc}
V.~Mukhanov, Physical Foundations of Cosmology, Cambridge University Press, Oxford (2005).




\bibitem{Weinberg:2008zzc}
S.~Weinberg, Cosmology, Oxford University Press (2008).




\bibitem{Choi:2025xye}
D.~Choi, S.~Jeon and J.~O.~Gong,
JCAP \textbf{05}, 006 (2026)
[arXiv:2509.21850 [astro-ph.CO]].




\bibitem{Ananda:2006af}
K.~N.~Ananda, C.~Clarkson and D.~Wands,
Phys. Rev. D \textbf{75}, 123518 (2007)
[arXiv:gr-qc/0612013 [gr-qc]].




\bibitem{Baumann:2007zm}
D.~Baumann, P.~J.~Steinhardt, K.~Takahashi and K.~Ichiki,
Phys. Rev. D \textbf{76}, 084019 (2007)
[arXiv:hep-th/0703290 [hep-th]].




\bibitem{Zeldovich:1967lct}
Y.~B.~Zel'dovich and I.~D.~Novikov,
Sov. Astron. \textbf{10}, 602 (1967).




\bibitem{Hawking:1971ei}
S.~Hawking,
Mon. Not. Roy. Astron. Soc. \textbf{152}, 75 (1971).




\bibitem{Carr:1974nx}
B.~J.~Carr and S.~W.~Hawking,
Mon. Not. Roy. Astron. Soc. \textbf{168}, 399-415 (1974).




\bibitem{Burgess:2007pt}
C.~P.~Burgess,
Ann. Rev. Nucl. Part. Sci. \textbf{57}, 329-362 (2007)
[arXiv:hep-th/0701053 [hep-th]].




\bibitem{Nakayama:2008wy}
K.~Nakayama, S.~Saito, Y.~Suwa and J.~Yokoyama,
JCAP \textbf{06}, 020 (2008)
[arXiv:0804.1827 [astro-ph]].




\bibitem{Choi:2024ilx}
G.~Choi, W.~Ke and K.~A.~Olive,
Phys. Rev. D \textbf{109}, no.8, 083516 (2024)
[arXiv:2402.04310 [hep-ph]].




\bibitem{Khlebnikov:1997di}
S.~Y.~Khlebnikov and I.~I.~Tkachev,
Phys. Rev. D \textbf{56}, 653-660 (1997)
[arXiv:hep-ph/9701423 [hep-ph]].




\bibitem{Kohri:2018awv}
K.~Kohri and T.~Terada,
Phys. Rev. D \textbf{97}, no.12, 123532 (2018)
[arXiv:1804.08577 [gr-qc]].




\bibitem{Inomata:2019ivs}
K.~Inomata, K.~Kohri, T.~Nakama and T.~Terada,
Phys. Rev. D \textbf{100}, 043532 (2019)
[erratum: Phys. Rev. D \textbf{108}, no.4, 049901 (2023)]
[arXiv:1904.12879 [astro-ph.CO]].




\bibitem{Harada:2013epa}
T.~Harada, C.~M.~Yoo and K.~Kohri,
Phys. Rev. D \textbf{88}, no.8, 084051 (2013)
[erratum: Phys. Rev. D \textbf{89}, no.2, 029903 (2014)]
[arXiv:1309.4201 [astro-ph.CO]].




\bibitem{Escriva:2019phb}
A.~Escriv{\`a}, C.~Germani and R.~K.~Sheth,
Phys. Rev. D \textbf{101}, no.4, 044022 (2020)
[arXiv:1907.13311 [gr-qc]].




\bibitem{Yoo:2019pma}
C.~M.~Yoo, J.~O.~Gong and S.~Yokoyama,
JCAP \textbf{09}, 033 (2019)
[arXiv:1906.06790 [astro-ph.CO]].




\bibitem{Yoo:2020dkz}
C.~M.~Yoo, T.~Harada, S.~Hirano and K.~Kohri,
PTEP \textbf{2021}, no.1, 013E02 (2021)
[erratum: PTEP \textbf{2024}, no.4, 049203 (2024)]
[arXiv:2008.02425 [astro-ph.CO]].




\end{thebibliography}
\end{document}